\documentclass[prd,twocolumn,showpcs,amsmath,amssymb,nofootinbib,preprintnumbers,balancelastpage]{revtex4}
\pdfoutput=1
\usepackage{epsfig}
\usepackage{amsmath}
\usepackage{bm}
\usepackage{times}
\usepackage{graphicx}
\usepackage{color}
\usepackage{slashed}
\usepackage{graphicx}
\usepackage{amsmath, amssymb}
\usepackage{xfrac}

\def\bea{\begin{eqnarray}}
\def\eea{\end{eqnarray}}
\def\ba{\begin{eqnarray}}
\def\ea{\end{eqnarray}}
\def\be{\begin{equation}}
\def\ee{\end{equation}}

\begin{document}
\preprint{CALT 68-2880}

\title{Particle creation in a toroidal universe}

\author{Bartosz Fornal\\
\textit{California Institute of Technology, Pasadena, CA 91125, USA}\\
}
\date{\today}

\begin{abstract}
We calculate the particle production rate in an expanding universe with a three-torus topology. We discuss also the complete evolution of the size of such a universe. The energy density of particles created through the nonzero modes is computed for selected masses. The unique contribution of the zero mode and its properties are also analyzed.
\vspace{8mm}
\end{abstract}

\maketitle
\bigskip

\section{Introduction}
Although current astrophysical observations provide precise information on the geometry of the universe \cite{Komatsu:2010fb}, its topology remains a mystery. We don't even know whether the universe is compact or infinite. Nevertheless, lower bounds can be put on its size for each compact topology (see, \cite{Glen} and references therein).

Amongst the possible topologies for the universe those with some or all spatial dimensions compactified are especially interesting, since then Casimir energies provide an additional contribution to the energy density. This may lead to an interesting  vacuum structure for the standard model coupled to gravity which is insensitive to quantum gravity effects \cite{ArkaniHamed:2007gg, AFW, FW}. The simplest flat topology with all dimensions compactified is a three-torus, and that is the topology we will concentrate on throughout this paper.

The scenario of our universe having a three-torus topology was investigated by many authors (see, \cite{Linde} and references therein). Probably one of the most appealing features of such a model is that the creation of a three-torus universe is much more likely to occur than that of an infinite flat or closed universe \cite{Zeldov,Linde}. In addition, it has also been shown that a three-torus topology can provide convenient initial conditions for inflation \cite{Zeldov}.

Here we consider gravitational particle creation in an expanding toroidal universe. The particle production formalism was developed in \cite{Parker1,Parker2,Parker3} and investigated in great detail in later works (see, \cite{Zeldovich,Starob,Grib} and references therein). Since then, it has been thoroughly studied in the case of a FRW cosmology, including its implications for dark matter creation around the inflationary epoch \cite{Chung}.  However, particle production in a toroidal universe hasn't been extensively studied (see, \cite{Berger} for some work on the subject).

In this paper we provide a detailed numerical calculation of the particle production in a three-torus universe. We start with introducing the relevant formalism. We then discuss the evolution of the size and energy density of the universe from the Planck time to the present time. Next, we derive analytical formulae for the particle number and energy density at early and late times. We then find full numerical solutions and confirm that they agree with the analytical approximations in the appropriate regions. The particle production through the nonzero modes is somewhat similar as in the closed universe case discussed in \cite{Starob}. However, the three-torus particle creation includes an additional contribution from the zero mode, which strongly depends on the choice of initial conditions.

\section{Three-torus metric}
We start with the following spacetime interval,
\bea
ds^2 = - d t^2 + t_{i j} d y^i d y^j\,,
\eea
where $t_{i j}$ is the metric on the three-torus with $i, j = 1, 2, 3$, and the compact coordinates are $y^i \in [0, 2\pi)$.
The $3\times3$ matrix with components $t_{ij}$ is positive definite and has a determinant equal to the volume modulus $a^3$. A  suitable parametrization is given by,
\bea\label{torus1}
t_{i j} = \frac{a^2}{(\rho_3 \tau_2)^{2/3}}\left(
                              \begin{array}{ccc}
                                1 & \tau_1 & \rho_1 \\
                                \tau_1 & \tau_1^2+\tau_2^2 & \rho_1\tau_1 + \rho_2\tau_2 \\
                               \rho_1 & \rho_1\tau_1+ \rho_2\tau_2 & \rho_1^2+\rho_2^2+\rho_3^2 \\
                              \end{array}
                            \right) ,
\eea
where $(\tau_1, \tau_2, \rho_1, \rho_2, \rho_3)$ are the shape moduli. We assume that all the parameters in (\ref{torus1}) are independent of the spatial coordinates. Furthermore, we seek stable solutions to Einstein's equations for which the shape parameters are also constant in time.  It was shown in \cite{FW} that this is only possible for,
\bea
\label{shape}
(\tau_1, \tau_2, \rho_1, \rho_2, \rho_3)= \left(\frac{1}{2}, \frac{\sqrt{3}}{2}, \frac{1}{2}, \frac{\sqrt{3}}{6}, \frac{\sqrt{6}}{3}\right),
\eea
which arises from the symmetries of the Casimir energies. For a three-torus universe characterized by the parameters (\ref{shape})  only the evolution of the volume modulus is nontrivial. The corresponding metric takes the form,
\bea\label{torus}
t_{i j} = \frac{a^2}{\sqrt[3]{4}}\left(
                              \begin{array}{ccc}
                                2 & 1 & 1 \\
                                1 & 2 & 1 \\
                                1 & 1 & 2 \\
                              \end{array}
                            \right) ,
\eea
and this is the metric we will adopt in our further analysis.

\section{Particle production}
We first derive general formulae for the gravitational particle production rate in a three-torus universe.
Let us consider a complex scalar field $\Psi=\Psi(x)$ of mass $m$ with the Lagrangian density given by,
\bea
\mathcal{L} = \sqrt{-g} \left[g^{\mu \nu}\partial_{\mu} \Psi \,\partial_\nu \Psi^* - \left(m^2+\frac{R}{6}\right) |\Psi|^2\right]\ ,
\eea
where $R$ is the Ricci scalar,
\bea
R = \frac{6}{a^2} \left(a\,\ddot{a}+\dot{a}^2\right)\ .
\eea
The dot denotes the derivative with respect to time $t$ and the factor $\xi=1/6$ was chosen to have conformal invariance in the limit $m\rightarrow0$.
We note, however, that the results in this paper are not very sensitive to this choice and would be similar, for example, in the case of a minimally coupled scalar field (for which $\xi=0$).

The stress-energy tensor is given by,
\bea\label{tensor}
T_{\mu\nu} &=& \partial_{\mu} \Psi \,\partial_\nu \Psi^* + \partial_{\nu} \Psi \,\partial_\mu \Psi^* - g_{\mu\nu}\frac{\mathcal{L}}{\sqrt{-g}}\nonumber\\
& &  -\, \frac{1}{3}\left(R_{\mu\nu} + \nabla_\mu \nabla_\nu - g_{\mu\nu}\nabla_\gamma \nabla^\gamma \right) |\Psi|^2\ .
\eea
The equation of motion for the field $\Psi$ is,
\bea\label{eq}
\ddot\Psi + 3 \frac{\dot{a}}{a}\dot\Psi - \nabla^2\Psi + \left(m^2+\frac{\ddot{a}}{a}+\frac{\dot{a}^2}{a^2}\right)\Psi = 0\ ,
\eea
where $\nabla^2$ is the Laplacian on the three-torus.
We write the solutions of equation (\ref{eq}) as,
\bea
\Psi_\lambda(x) = u_\lambda(t) \,\phi_\lambda(\vec{y})\ ,
\eea
with $\lambda = (\lambda_1, \lambda_2, \lambda_3)$.
In our case,
\bea\label{nabla}
\nabla^2\phi_\lambda = - \frac{\tilde{t}_{i j}}{a^2} \lambda_i \lambda_j \phi_\lambda,
\eea
where,
\bea
\tilde{t}_{ij} \equiv t_{ij}/a^2\ ,
\eea
and there is an implicit sum over the repeated indices $i, j = 1, 2, 3$, with $\lambda_i = 0, \pm1, \pm2, ...$. The formula for $\phi_\lambda$
is given by,
\bea
\phi_\lambda = C\, e^{i \,\tilde{t}_{k l} \lambda_k y^l}\ .
\eea
Note that for $\lambda = (0, 0, 0)$ we have $\phi_0 = \rm const$, which corresponds to the zero mode.
Equation (\ref{eq}) takes the form,
\bea\label{eq2}
\ddot u_\lambda + 3 \frac{\dot{a}}{a}\dot u_\lambda + \left(\frac{\omega_\lambda^2}{a^2}+\frac{\ddot{a}}{a}+\frac{\dot{a}^2}{a^2}\right) u_\lambda = 0\ ,
\eea
where,
\bea\label{omega}
\omega_\lambda^2 &=&  (m a)^2  +  \tilde{t}_{i j} \lambda_i \lambda_j\ .
\eea
We can now quantize the field introducing standard canonical equal-time commutation relations for the field and its generalized momentum. Those relations are satisfied if we write the field $\Psi$ as,
\bea
\hspace{-2mm}\hat{\Psi} = \frac{1}{(2\pi a)^{\sfrac{3}{2}}} \sum_{\lambda_1, \lambda_2, \lambda_3 = -\infty}^\infty \left[\phi_\lambda \,u^*_\lambda \,\hat{a}_\lambda +  \phi^*_\lambda \,u_\lambda  \,\hat{b}_\lambda^\dagger\right],
\eea
where $\hat{a}^\dagger_\lambda$ and $\hat{a}_\lambda$ are the creation and annihilation operators of a particle in the state $\lambda = (\lambda_1, \lambda_2, \lambda_3)$, and $\hat{b}^\dagger_\lambda$ and $\hat{b}_\lambda$ are those for antiparticles, all obeying the usual commutation relations.
Adopting such a convention, the Hamiltonian can be written as \cite{Starob},
\bea\label{ham}
\hspace{-10mm}\hat{H} &=& \hspace{-3mm} \sum_{\lambda_1, \lambda_2, \lambda_3 = -\infty}^\infty \omega_\lambda \bigg[A_\lambda\left(\hat{a}_\lambda \,\hat{a}_\lambda^{\dagger} + \hat{b}_{\bar{\lambda}}^{\dagger} \,\hat{b}_{\bar{\lambda}}\right)\nonumber\\
& & \hspace{22mm} +\, B_\lambda\, \hat{a}_\lambda^{\dagger} \,\hat{b}_{\bar{\lambda}}^{\dagger} + B^*_\lambda\, \hat{a}_{\lambda}\, \hat{b}_{\bar{\lambda}} \bigg],
\eea
with,
\vspace{-3mm}
\bea\label{omega2}
A_\lambda &=& \frac{a^2 |\dot{u}_\lambda|^2}{2\,\omega_\lambda}+\frac{1}{2}\omega_\lambda |u_\lambda|^2\ ,\nonumber\\
B_\lambda &=& \frac{a^2 \dot{u}_\lambda^2}{2\,\omega_\lambda}+\frac{1}{2}\omega_\lambda u_\lambda^2\ ,
\eea
and ${\bar{\lambda}}$ defined through $\phi_{\bar{\lambda}} =  \phi_{\lambda}^*$.

In general, the Hamiltonian (\ref{ham}) is non-diagonal. The requirement of it being diagonal at some initial time $t_0$ imposes the following conditions,
\bea
u_\lambda(t_0) = \frac{1}{\sqrt{\omega_\lambda(t_0)}}\ , \ \ \ \dot{u}_\lambda(t_0) = i \,\frac{\sqrt{\omega_\lambda(t_0)}}{a(t_0)}\ .
\eea
Now, we can diagonalize the Hamiltonian through the following Bogoliubov transformation,
\bea
\hat{a}_\lambda &=& \alpha^*_\lambda(t)\, \hat{a}_\lambda'(t) + \beta_\lambda(t) \,\hat{b}_{\bar{\lambda}}'^{\dagger}(t) \ ,\nonumber\\
\hat{b}_\lambda &=& \alpha^*_\lambda(t) \,\hat{b}_\lambda'(t) + \beta_\lambda(t) \,\hat{a}_{\bar{\lambda}}'^{\dagger}(t) \ ,
\eea
where $|\alpha_\lambda|^2 - |\beta_\lambda|^2 = 1$.
It can be shown \cite{Starob}, from the requirement that there be no non-diagonal terms $\hat{a}_\lambda'^{\dagger} \hat{b}_\lambda'^{\dagger}$  or $\hat{a}_\lambda' \hat{b}_\lambda'$, that the equations for $\alpha_\lambda(t)$ and $\beta_\lambda(t)$ are,
\bea
\dot{\beta}_\lambda &=& \frac{\dot{\omega}_\lambda}{2\,\omega_\lambda} \,\alpha_\lambda \exp\left(-2 i \int_{t_0}^t \frac{\omega_\lambda(t')}{a(t')} dt'\right)\ , \label{coupled}\\
\dot{\alpha}_\lambda &=& \frac{\dot{\omega}_\lambda}{2\,\omega_\lambda}\, \beta_\lambda \exp\left(2 i \int_{t_0}^t \frac{\omega_\lambda(t')}{a(t')} dt'\right)\ ,\label{coupled2}
\eea
with the initial conditions $\beta_\lambda(t_0)=0$ and $\alpha_\lambda(t_0)=1$.
The function $u_\lambda$ is expressed in terms of $\alpha_\lambda$ and $\beta_\lambda$ as \cite{Starob},
\bea
\hspace{-10mm} u_\lambda &=& \frac{1}{\sqrt{\omega_\lambda}} \bigg[\alpha^*_\lambda\exp\left(i \int_{t_0}^t \frac{\omega_\lambda(t')}{a(t')} dt'\right)\nonumber\\
& & \hspace{15mm}+\, \beta_\lambda\exp\left(-i \int_{t_0}^t \frac{\omega_\lambda(t')}{a(t')} dt'\right)\bigg]\ .
\eea
\begin{table*}[t]
\begin{center}
\begin{tabular}[t]{|c|c|c|c|c|c|}
  \hline
\multicolumn{1}{|c|}{} & \multicolumn{5}{|c|}{epoch} \\ \cline{2-6}
  & \,\,\,pre-inflationary era\,\, & \,\,\,inflation \,\,  & \,\,\,radiation era\,\,  & \ \ \,\,matter era\, \  \ & dark energy era\\
    \hline\hline
         & $\approx 5\cdot 10^{-44} \ \rm s$  &\,  $\approx 5\cdot10^{-36} \ \rm s $   \,&\, $\approx 10^{-33} \ \rm s $   \,& \, $\approx 7\cdot10^4 \ \rm years$   \, & $\approx 10^{10} \ \rm years$\\
\ \raisebox{1.8ex}[0pt]{$t_{\rm init}$}  \    &    $\simeq 8\cdot 10^{-20} \ \rm GeV^{-1}$    &  $\simeq 7\cdot10^{-12} \ \rm GeV^{-1}$    &  $\simeq 10^{-9} \ \rm GeV^{-1}$   &  $\simeq 3\cdot 10^{36} \ \rm GeV^{-1}$  &   $\simeq 5\cdot 10^{41} \ \rm GeV^{-1}$\\ \hline
       \ \    & \, $\approx 5\cdot10^{-36} \ \rm s $   \,&  \, $\approx 10^{-33} \ \rm s $    \,& \, $\approx 7\cdot10^4 \ \rm years$   \,&  $\approx 10^{10} \ \rm years$  &    $\approx 1.37\cdot10^{10} \ \rm years$\\
\ \raisebox{1.8ex}[0pt]{$t_{\rm final}$}  \      & \, $\simeq 7\cdot10^{-12} \ \rm GeV^{-1}$   \, & \,  $\simeq 10^{-9} \ \rm GeV^{-1}$   \, & \, $\simeq 3\cdot 10^{36} \ \rm GeV^{-1}$   \, & \, $\simeq 5\times 10^{41} \ \rm GeV^{-1}$   \, &  $\simeq 7\cdot 10^{41} \ \rm GeV^{-1}$\\ \hline
       \ \   &    \,  &    &\,   \,& \,    \, & \\
       \ \   &  \raisebox{1.8ex}[0pt]{$a_{p0} \sqrt{t}$}    & \raisebox{1.8ex}[0pt]{$a_{i0}\exp\left[\sqrt{\rho_{\rm tot}/(3M_p^2)} \,t\right]$}&\ \raisebox{1.8ex}[0pt]{$a_{r0} \sqrt{t}$}   \,& \,   \raisebox{1.8ex}[0pt]{$a_{m0} t^{\sfrac{2}{3}}$}  \, & \raisebox{1.8ex}[0pt]{$a_{d0}\exp\left[\sqrt{\rho_{\rm tot}/(3M_p^2)} \,t\right]$}\\
\ \raisebox{3.5ex}[0pt]{$a(t)$}  \     &  \,\raisebox{1.2ex}[0pt]{$a_{p0} \simeq 4\cdot10^{-10} \frac{1}{\sqrt{\rm GeV}}$}    &  \raisebox{1.2ex}[0pt]{$a_{i 0} \simeq 10^{-15} \ \rm GeV^{-1}$} & \, \raisebox{1.2ex}[0pt]{$a_{r0} \simeq 6\cdot10^{20} \frac{1}{\sqrt{\rm GeV}}$} & \, \raisebox{1.2ex}[0pt]{$a_{m 0} \simeq 6\cdot10^{14} \frac{1}{\sqrt[3]{\rm GeV}}$}   & \raisebox{1.2ex}[0pt]{$a_{d 0} \simeq 2\cdot 10^{42} \ \rm GeV^{-1}$} \\ \hline
       \ \   &   $\approx 2\cdot 10^{-35} \ \rm m$    & $\approx 2\cdot 10^{-31} \ \rm m$   & $\approx 4 \ \rm m$   & \, $\approx 2\cdot 10^{23} \ \rm m$   \, & $\approx 8\cdot10^{26} \ \rm m$\\
\ \raisebox{1.8ex}[0pt]{$a(t_{\rm init})$}  \     & $\simeq 10^{-19} \ \rm GeV^{-1}$&  $ \simeq 10^{-15} \ \rm GeV^{-1}$    &  $\simeq  2\cdot 10^{16} \ \rm GeV^{-1}$   & \, $\simeq   10^{39}\ \rm GeV^{-1}$   \, & $\simeq 4 \cdot 10^{42} \ \rm GeV^{-1}$\\ \hline
       \ \      &  $\approx 2\cdot 10^{-31} \ \rm m$ &  \, $\approx 4 \ \rm m$   \,& \,$\approx 2\cdot 10^{23} \ \rm m$  \,& \, $\approx 8\cdot 10^{26} \ \rm m$   & $\approx 10^{27} \ \rm m$ \\
\ \raisebox{1.8ex}[0pt]{$a(t_{\rm final})$}  \      &  $ \simeq  10^{-15} \ \rm GeV^{-1}$ & \, $\simeq  2\cdot 10^{16} \ \rm GeV^{-1}$   \, & \, $\simeq  10^{39}\ \rm GeV^{-1}$   \, & \, $\simeq4\cdot 10^{42} \ \rm GeV^{-1}$   \, & $\simeq 5 \cdot 10^{42} \ \rm GeV^{-1}$\\ \hline
       \ \      & \,    \,&  \,    \,& \,   \,& \,    \, & \\
\ \raisebox{1.7ex}[0pt]{$\rho_{\rm tot}(t)$}  \      & \,  \raisebox{1.7ex}[0pt]{$3 M_p^2 / (4 t^2)$}   \, & \,  \raisebox{1.7ex}[0pt]{$\rm const$}   \, & \, \raisebox{1.7ex}[0pt]{$3 M_p^2 / (4 t^2)$}  \, & \, \raisebox{1.7ex}[0pt]{$4 M_p^2 / (3 t^2)$}   \, & \raisebox{1.7ex}[0pt]{$ \rm const$} \\ \hline
&  &  &  &  & \\
\raisebox{1.7ex}[0pt]{$\rho_{\rm tot}(t_{\rm init})$}      &  \raisebox{1.7ex}[0pt]{$\approx 7\cdot 10^{74} \ {\rm GeV^4}$}    & \raisebox{1.7ex}[0pt]{$\approx 9\cdot10^{58} \ \rm GeV^4$}    &  \raisebox{1.7ex}[0pt]{$\approx  4\cdot 10^{54} \ \rm GeV^4$}   & \raisebox{1.7ex}[0pt]{$\approx 8\cdot 10^{-37} \ \rm GeV^4$}   & \raisebox{1.7ex}[0pt]{$\approx 3\cdot10^{-47} \ \rm GeV^4$}\\ \hline
&  &  $\approx 9\cdot10^{58} \ \rm GeV^4$  &  &  & \\
$\rho_{\rm tot}(t_{\rm final})$       &  $\approx  9\cdot10^{58}\ \rm GeV^4$    & after reheating:  & \, $\approx 5\cdot 10^{-37} \ \rm GeV^4$  \, & \, $\approx 3\cdot10^{-47} \ \rm GeV^4$  \, & $\approx 3\cdot10^{-47} \ \rm GeV^4$\\
&  & $\approx  4\cdot 10^{54} \ \rm GeV^4$ &  &  & \\ \hline
\end{tabular}
\end{center}
\vspace{0mm}
\caption{\footnotesize{Estimated timeline for the evolution of a three-torus universe from the Planck time until the present time $t_{\rm univ} \approx 13.7 \ \rm billion \ years$, including the size of the universe and total energy density during different epochs, assuming a pre-inflationary expansion $a(t)=a_{p0}\sqrt{t}\,$ and the current size of the universe $a(t_{\rm univ}) \approx 10 \, R_H$.}}
\end{table*}
\hspace{-1mm}Relations (\ref{coupled}) and (\ref{coupled2}) can be combined into one second order differential equation for $\beta_\lambda(t)$,
\bea\label{Bog}
\ddot{\beta}_\lambda + \left(\frac{\dot{\omega}_\lambda}{\omega_\lambda} - \frac{\ddot{\omega}_\lambda}{\dot{\omega}_\lambda} +\frac{2 i \omega_\lambda}{a}\right)\dot{\beta}_\lambda - \frac{\dot{\omega}_\lambda^2}{4\,\omega_\lambda^2}\beta_\lambda = 0\ ,
\eea
with the initial conditions,
\bea\label{initiial}
\beta_\lambda(t_0) = 0\ , \ \ \ \dot{\beta}_\lambda(t_0) = \frac{\dot{\omega}_\lambda(t_0)}{2\,\omega_\lambda(t_0)}\ .
\eea
The normal ordered Hamiltonian now takes the diagonal form,
\bea\label{ham2}
\hat{H} &=& \hspace{-3mm}\sum_{\lambda_1, \lambda_2, \lambda_3 = -\infty}^\infty \omega_\lambda \left(\hat{a}_\lambda'^{\dagger} \,\hat{a}_\lambda' + \hat{b}_{\lambda}'^{\dagger} \,\hat{b}_{\lambda}'\right),
\eea
and the physical vacuum depends on time through,
\bea
\hat{a}_\lambda'(t) |0(t)\rangle = \hat{b}_\lambda'(t) |0(t)\rangle = 0\ .
\eea
It is now straightforward to arrive at the formula for the number density of particle pairs created after time $t$,
\bea\label{number}
n(t) = \frac{1}{(2\pi a)^3} \sum_{\lambda_1, \lambda_2, \lambda_3 = -\infty}^\infty |\beta_{\lambda}|^2\ .
\eea
The energy density and pressure of the created particles are calculated from the vacuum expectation values of the appropriate components of the stress-energy tensor (\ref{tensor}). After normal ordering the result is (compare with \cite{Starob}),
\bea\label{density}
\rho(t) &=& \frac{1}{4\pi^3 a^4} \sum_{\lambda_1, \lambda_2, \lambda_3 = -\infty}^\infty \omega_\lambda |\beta_{\lambda}|^2\ ,\\
p_{ii}(t) &=& \frac{\sqrt[3]{2}}{12\pi^3 a^4} \sum_{\lambda_1, \lambda_2, \lambda_3 = -\infty}^\infty \omega_\lambda \nonumber\\ \label{density2}
& & \times \left[|\beta_{\lambda}|^2-\frac{m^2 a^2}{2\,\omega_\lambda}\left(|u_\lambda|^2 - \frac{1}{\omega_\lambda}\right)\right],\\
p_{ij}(t) &=& \frac{p_{ii}(t)}{2} \hspace{4mm} {\rm for} \hspace{4mm} i\ne j\ , \label{density3}
\eea
where $i, j = 1, 2, 3$.
Note that the pressure is not isotropic in our case. It would be isotropic only for the simplest choice of the three-torus metric ${\rm diag}(a^2, a^2, a^2)$.

\section{Evolution of the three-torus universe}
In order to calculate the particle production rate, we first have to know how $a(t)$ evolved in time. Our case is  different from the usual FRW cosmology, in which $a(t)$ is the scale factor and can be rescaled by an arbitrary number. For a three-torus universe $a(t)$ has a physical meaning -- it describes the size of the universe at a given time $t$. Recent analyzes (see, \cite{Aneesh} and references therein) set a lower bound on its present value of $a(t_{\rm univ}) \gtrsim 6 R_H$, where $R_H$ is the Hubble radius today. Let's therefore assume that the current size of the three-torus universe is,
\bea
a(t_{\rm univ}) = 10\, R_H \approx 10^{27} \ \rm m \ .
\eea

Table I shows the estimated timeline, the size of the universe and the total energy density during different epochs in the evolution of the universe.
For each epoch the total energy density is uniquely determined by the expansion rate through the Friedmann equation,
\bea
\left(\frac{\dot{a}}{a}\right)^2 = \frac{\rho_{\rm tot}}{3 M_p^2}\ ,
\eea
where the reduced Planck mass $M_p \simeq 2.4 \cdot 10^{18} \ \rm GeV$.
Using the formulae for $a(t)$ for different epochs, we can evolve $a(t_{\rm univ})$ back in time and find its value at any particular instance in the past.

Observational data suggest that we currently live in a dark energy dominated universe, where the energy density is roughly constant and the expansion is exponential. It was preceded (at $t \lesssim 10 \ {\rm billion \ years}$) by a matter dominated era with the size of the universe increasing like $t^{2/3}$. Before the matter era (at $t \lesssim 70000 \ {\rm years}$), radiation was driving the expansion like $\sqrt{t}$. It is believed that before the radiation epoch there was a brief period of inflation ($10^{-35} \ {\rm s} \lesssim t\lesssim 10^{-33} \ {\rm s}$), an exponential expansion resulting from a constant energy density. The inflationary energy density in table I was estimated adopting $72$ e-folds of inflation and assuming that it ended at $t \approx 10^{-33} \ {\rm s}$.

The expansion rate in the pre-inflationary era is unknown. However, in the case of a three-torus universe it might have a natural explanation. As mentioned in the introduction, a compact topology results in the appearance of Casimir energies of existing fields. The full formulae for the Casimir energies in a three-torus universe for an arbitrary set of shape moduli was derived in \cite{FW}. For a real scalar field with mass $m \ll 10^{15} \ {\rm GeV}$ the Casimir energy density before inflation for our choice of metric is,
\bea\label{CE0}
\lefteqn{ \ \rho_{\rm Cas}(a) \ =\  -\frac{2}{(2\pi)^4}\frac{1}{a^4}\, \Bigg\{ \ \frac{\sqrt{3}}{2^{2/3}}\frac{\pi}{12}+\frac{2^{1/3}}{3^{3/2}}\,\frac{\zeta(3)}{\pi}+ 2^{1/3}\frac{\pi^2}{360} }\nonumber\\
& & \hspace{-2mm}\!\!+\,\frac{2^{11/6}}{3^{3/4}}\!\!\!\sum_{n_2, n_3=1}^\infty  \left(\frac{n_3}{n_2}\right)^{3/2}\cos\left(\pi\,n_2\,n_3\right)
\,K_{3/2}\!\left(\sqrt{3}\,\pi \,n_2 \,n_3\right)\nonumber\\
& & \!\!\!\!\!\!+\!\!\!\!\!\sum_{n_2, n_3 = -\infty}^\infty\!\!\!\!\!\!\!'\ \ \ \sum_{n_1=1}^\infty \cos\!\Big[\tfrac{2}{3}\pi n_1 (n_2\! +\!n_3)\Big]\sqrt{\tfrac{2^{5/3}}{3}(n_2^2\!-\!n_2 n_3\! +\!n_3^2)}\nonumber\\
& & \ \ \, \times \, \frac{1}{n_1}\,K_1\!\left(\tfrac{4\sqrt{2}}{3}\pi n_1\sqrt{(n_2^2-n_2 n_3 +n_3^2)}\right)\Bigg\}\ ,
\eea
which can be written as,
\bea\label{Cas}
\rho_{\rm Cas}(a) =  -\frac{\kappa}{a^4}\ ,
\eea
where $\kappa\simeq 5 \times 10^{-3}$. In a general case, this formula includes also a factor corresponding to the number of degrees of freedom for a given field and an additional minus sign for fermions. Interestingly, the Casimir energies at $a \ll 1/m$ satisfy the same equation of state as radiation, i.e.,
\bea
p_{\rm Cas}(a(t)) = \frac{1}{3}\, \rho_{\rm Cas}(a(t))\ .
\eea
Therefore, if there were more fermionic degrees of freedom ($n_f$) than bosonic ($n_b$) and Casimir energies dominated the total energy density before inflation, the universe would expand according to $a(t) = a_{0} \sqrt{t}$, with the total energy density given by,
\bea
\rho^{\rm tot}_{\rm Cas}(a) = \frac{(n_f-n_b)\kappa}{a^4}\ .
\eea
Surprisingly, if we choose in our case $(n_f-n_b)\approx 20$, we obtain the total Casimir energy density at the time when inflation started,
\bea\label{condition2}
\rho^{\rm tot}_{\rm Cas}(a(t_{\rm inf})) \approx \rho_{\rm tot}(t_{\rm inf})\ ,
\eea
where $a(t_{\rm inf})$ is the universe size at the beginning of inflation and $\rho_{\rm tot}(t_{\rm inf})$ is the total energy density of the universe during inflation,
both of which where estimated before independently of the Casimir energies through the evolution back in time. Of course, the required value of $(n_f-n_b)$ which satisfies condition (\ref{condition2}) strongly depends on the inflationary parameters.

\begin{figure}[t]
\centerline{\scalebox{1.45}{\includegraphics{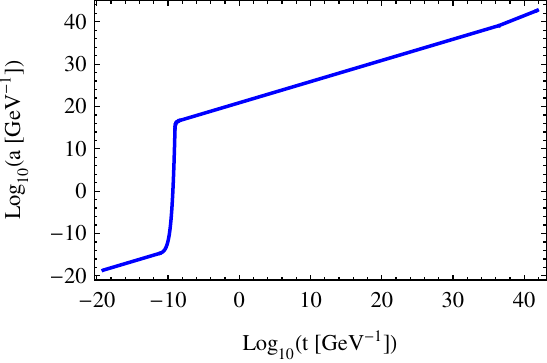}}}
\caption{\footnotesize{Size of the three-torus universe as a function of time, assuming $a(t_{\rm univ}) = 10\, R_H$.}}
\end{figure}
\begin{figure}[t]
\centerline{\scalebox{1.45}{\includegraphics{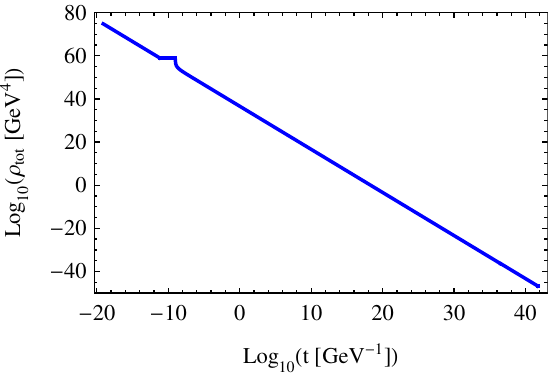}}}
\caption{\footnotesize{Total energy density in the three-torus universe as a function of time.}}
\end{figure}

Another interesting observation concerning the three-torus topology is that assuming the current size of the universe $a(t_{\rm univ}) \approx 10\,R_H$ and a pre-inflationary expansion $a(t) = a_{p0} \sqrt{t}$, the evolution back in time yields the size at Planck time $a(t_p) \approx  l_p$, where $l_p$ is the Planck length.
Note that an evolution from the Planck size at Planck time to ten Hubble radii at present time wouldn't be possible if we assumed a universe with a closed topology, since then the curvature term contribution would entirely dominate the total energy density during the early stages of the universe evolution.

The plot of the size of the universe $a(t)$ corresponding to the values from table I is given in figure 1. A similar plot for the total energy density is shown in figure 2.
We note that a similar calculation to the one presented in this paper can be done assuming a different pre-inflationary expansion.

Knowing the shape of $a(t)$, we can now numerically solve equation (\ref{Bog}) for the Bogoliubov coefficients and then use formulae (\ref{number}) and (\ref{density}) to calculate the number density and energy density  for the gravitationally created pairs of scalar particles of mass $m$. Our results can be easily generalized for other types of particles by adopting the appropriate form of the stress-energy tensor. We will perform the calculation for three different masses: $m_1=10^{9} \ \rm GeV$, $m_2=10^3 \ \rm GeV$, and $m_3=10^{-3} \ \rm GeV$, without the zero mode first, and then computing its contribution as well. In our calculation we will be using natural units, i.e.,
\bea
 2\cdot10^{-16}\ \rm m \ \approx  \  1 \ {\rm GeV^{-1}} \ \approx \ 7\cdot 10^{-25}\ \rm s\ .
\eea
We will also assume that there is no back-reaction of the created particles on the background evolution.

\section{Analytical results for nonzero modes}
In order to fully understand the numerical solutions it is very helpful to derive their analytical behavior in two regions: $t \ll 1/m$ and $t \gg 1/m$. In contrast to \cite{Zeldovich,Starob}, we choose not to specify the initial conditions at the singularity, but at the Planck time, since the physics at earlier times is unknown,
\bea\label{init}
\beta_\lambda(t_p) = 0\ , \ \ \ \dot{\beta}_\lambda(t_p) = \frac{\dot{\omega}_\lambda(t_p)}{2\,\omega_\lambda(t_p)}\ .
\eea
However, we note that for the nonzero modes the solutions are insensitive to the value of $t_0$ at which we impose those conditions, as long as we look at the region $t \gg t_0$.

Let us first consider the case $m t \ll 1$ and focus only on the pre-inflationary epoch, for which we adopted $a(t) = a_{p0} \sqrt{t}$.
With the additional assumption $m a \ll 1$ the largest contribution to the sum (\ref{number}) comes from terms with small $\lambda_i$. Thus, we can assume $\lambda_i t \ll a$, in which case the solution to equation (\ref{Bog}) is given by,
\bea\label{nn}
\beta_{\lambda}(t) \approx \frac{\sqrt{\omega_\lambda}}{2 (\tilde{t}_{ij} \lambda_i \lambda_j)^{\sfrac{1}{4}}}- \frac{(\tilde{t}_{ij} \lambda_i \lambda_j)^{\sfrac{1}{4}}}{2\,\sqrt{\omega_\lambda}}\ .
\eea
After plugging (\ref{nn}) into (\ref{number}), the particle number density is,
\bea\label{nden}
\hspace{-10mm}n(t) &\approx& \frac{1}{(2\pi a)^3} \sum_{\lambda_1, \lambda_2, \lambda_3 = -\infty}^\infty\!\!\!\!\!\!\!\!\!\!'  \ \ \ \  \frac{\left[\omega_\lambda - (\tilde{t}_{ij}\lambda_i\lambda_j)^{\sfrac{1}{2}}\right]^2}{4\, \omega_\lambda (\tilde{t}_{ij}\lambda_i\lambda_j)^{\sfrac{1}{2}}}\nonumber\\
&\approx & \frac{m^4 a}{128 \pi^3} \sum_{\lambda_1, \lambda_2, \lambda_3 = -\infty}^\infty\!\!\!\!\!\!\!\!\!\!' \ \ \ \ \ \  \frac{1}{(\tilde{t}_{ij}\lambda_i\lambda_j)^{2}}
\ \simeq \ \frac{m^4 a}{250}\ ,
\eea
where the prime excludes the zero mode.

The calculation of the energy density and pressure in the region $m t \ll 1$ cannot be performed with the same assumption $\lambda_i t \ll a$, as now terms with large $\lambda_i$ contribute significantly. However, for $m a \ll 1$ we have $|\beta_\lambda| \ll 1$ and $\alpha_\lambda \approx 1$. With equation (\ref{coupled}) we can approximate $\beta_\lambda(t)$ by,
\bea
\beta_\lambda &\approx& \frac{m^2}{2 \,\tilde{t}_{ij}\lambda_i\lambda_j}\int_0^t a(t') \,\dot{a}(t') \nonumber\\
& & \hspace{10mm}\times \exp\left(-2 i \frac{(\tilde{t}_{ij}\lambda_i\lambda_j)^{\sfrac{1}{2}}}{a(t')}\, t'\right) dt'\ .
\eea
Using this result to rewrite (\ref{density}), we arrive at the energy density for the created particles during the pre-inflationary era,
\bea\label{rrho}
\rho(t) &\approx& \frac{m^4 a_{p0}^4}{64 \pi^3 a^4}  \sum_{\lambda_1, \lambda_2, \lambda_3 = -\infty}^\infty\!\!\!\!\!\!\!' \ \ \ \  \frac{1}{(\tilde{t}_{ij}\lambda_i\lambda_j)^{\sfrac{3}{2}}}
 \int_0^t dt_1 \int_0^t d t_2 \nonumber\\
& & \times\exp\left[-2 i (\tilde{t}_{ij}\lambda_i\lambda_j)^{\sfrac{1}{2}}\frac{1}{a_{p0}}\left(\sqrt{t_1}-\sqrt{t_2}\right)\right]\ .
\eea
The biggest contribution in equation (\ref{rrho}) comes from summing the terms with large $\lambda_i$. It turns out that we can approximate our three-torus by a three-sphere of radius $\sqrt[6]{2}\, a$, for which it is possible to perform the sum over $\lambda$'s. Equation (\ref{rrho}) takes the form,
\bea\label{rrho2}
\hspace{5mm}\rho(t) &\approx& \frac{m^4 a_{p0}^4}{64 \sqrt[3]{2}\,\pi^3 a^4}   \int_0^t dt_1 \int_0^t d t_2\nonumber\\
& & \times \sum_{\lambda = 1}^\infty\frac{1}{\lambda}
\exp\left[-2 i \frac{\lambda}{a_{p0}}\left(\sqrt{t_1}-\sqrt{t_2}\right)\right]\nonumber\\
& \hspace{-25mm}= & \hspace{-14mm} \frac{m^4 a_{p0}^4}{64 \sqrt[3]{2}\,\pi^3 a^4}  \! \int_0^t \!\!\!dt_1  \!\!\int_0^t \!\!\! d t_2\log\!\!\left[\frac{1}{1\!- \exp\left[\frac{2 i}{a_{p0}}\!\left(\sqrt{t_1}\!-\!\sqrt{t_2}\right)\right]}\right].\nonumber\\
\eea
The double-integral above is real and positive. The resulting energy density for $m t \ll 1$ is essentially constant in time.

Using a similar approximation one can derive the formula for the pressure,
\bea
p_{ij}(t) &\approx& \tilde{t}_{ij}\frac{\rho(t)}{3} - \tilde{t}_{ij}\frac{m^4 a_{p0}^2}{96 \pi^3 a^2}  \Bigg\{ \frac{m^2 a_{p0}^2}{4} \int_0^t dt_1 \int_0^t d t_2\nonumber\\
& & \hspace{5mm}\times \,{\rm Li}_3\left(\exp\left[\frac{2 i}{a_{p0}}\left(\sqrt{t_1}-\sqrt{t_2}\right)\right]\right)\nonumber\\
& &\hspace{-16mm} - \, 2 \,{\rm Re}\int_0^t d t_1 \log\left[1- \exp\left(\frac{2 i}{a_{p0}}\left(\sqrt{t_1}-\sqrt{t}\right)\right)\right]
\Bigg\}.
\eea
A numerical check shows that for $m t \ll 1$ the pressure for the created particles is also almost constant and satisfies the ``quasi-vacuum-like'' equation of state, \bea
p_{ij}(t) \approx -\,\tilde{t}_{ij}\,\rho(t)\ .
\eea

We will not go into the details of calculating the particle number or energy density in a three-torus universe for $m t \gg 1$, since this case is analogous to that of an infinite flat or closed universe, which was explored in detail in \cite{Starob}. Briefly summarizing, for $m t \gg 1$ all Bogoliubov coefficients $\beta_\lambda$ are roughly constant and particle creation doesn't occur. For a universe expanding according to $a(t)=a_0 t^q$, where $0<q\leq 2/3$, the number density and the energy density of the created particles behave in the following way,
\bea\label{largen}
n(t) &\sim& \frac{m^{3-3q}}{t^{3q}} \sim \frac{1}{a^3}\ , \label{largen2}\\
\rho(t) &\sim& \frac{m^{4-3q}}{t^{3q}} \sim \frac{1}{a^3}\ ,
\eea
while the pressure \ $p_{ij}(t) \ll \rho(t)$.

\section{Numerical results for nonzero modes}
Having an analytical understanding of how the functions $n(t)$ and $\rho(t)$ behave for $m t \ll 1$ and $m t \gg 1$, in this section we discuss the full numerical solutions.

\begin{figure}[t]
\centerline{\scalebox{1.5}{\includegraphics{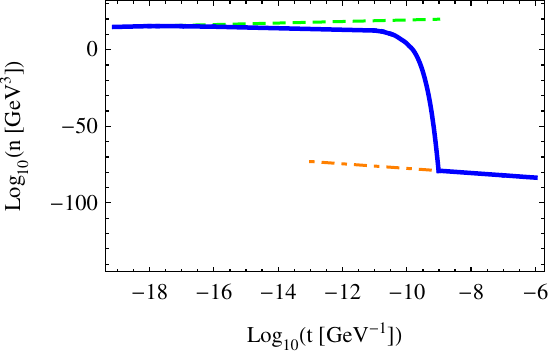}}}
\caption{\footnotesize{Number density of the created particles for $m_1=10^9 \ \rm GeV$  (blue solid line). The green dashed line corresponds to the $m t \ll 1$ and $m a \ll 1$ approximation given by equation (\ref{nden}) and the orange dot-dashed line corresponds to the $m t \gg 1$ approximation given by equation (\ref{largen}).}}
\end{figure}
\begin{figure}[t]
\centerline{\scalebox{1.5}{\includegraphics{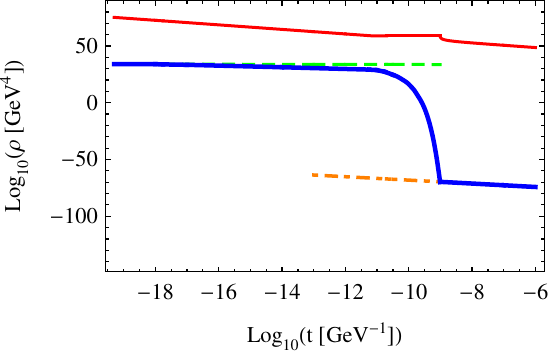}}}
\caption{\footnotesize{Energy density of the created particles for $m_1=10^9 \ \rm GeV$  (blue solid line). The green dashed line corresponds to the constant density for $m t \ll 1$ given by equation (\ref{rrho2}) and the orange dot-dashed line corresponds to the $m t \gg 1$ approximation given by equation (\ref{largen}). For comparison, the total energy density of the universe is plotted in solid red.}}
\end{figure}

We first solve numerically equation (\ref{Bog}) and find $\beta_\lambda(t)$  for different sets of $\lambda_i$'s, with $\omega_\lambda(t)$ defined through equation (\ref{omega}). The initial conditions are chosen according to (\ref{init}). We then plug the calculated $\beta_\lambda(t)$ to equations (\ref{number}) and (\ref{density}), and in this way obtain the particle number and energy density.

Figure 3 shows the plot of the particle number density (in blue) corresponding to a mass of the created particles of $m_1=10^9 \ \rm GeV$. The green dashed line increasing like $\sqrt{t}$ is the $m t \ll 1$ and $m a \ll 1$ approximation given by equation (\ref{nden}). Since the particle mass is large, this approximation breaks down relatively quickly and the number density starts slowly decreasing. During the inflationary epoch, it falls down rapidly by many orders of magnitude. Particles of mass $10^9 \ \rm GeV$ are no longer produced after inflation. This is confirmed by fitting the line falling off like $t^{-3/2}$ (orange dot-dashed line) corresponding to the relation (\ref{largen2}).

\begin{figure}[t]
\centerline{\scalebox{1.5}{\includegraphics{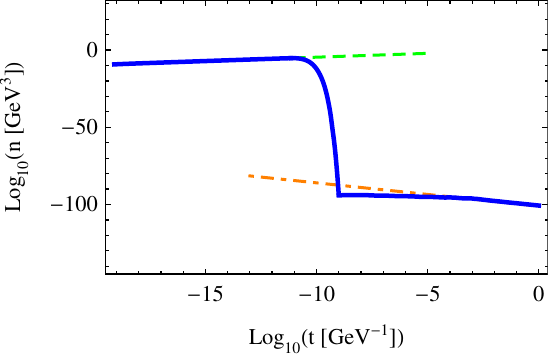}}}
\caption{\footnotesize{Same as figure 3, but for $m_2=10^3 \ \rm GeV$.}}
\end{figure}
\begin{figure}[t]
\centerline{\scalebox{1.5}{\includegraphics{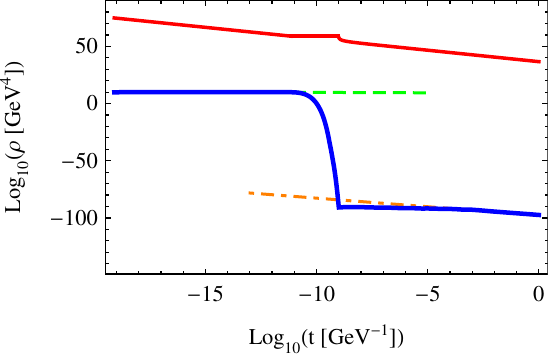}}}
\caption{\footnotesize{Same as figure 4, but for $m_2=10^3 \ \rm GeV$.}}
\end{figure}

Figure 4 shows the time evolution of the energy density for the created particles of mass $m_1=10^9 \ \rm GeV$. The energy density is nearly constant for $m t \ll 1$ and its value agrees with that from equation (\ref{rrho2}), represented by the green dashed line. The subsequent behavior of the energy density is similar as in the number density case -- following a mild decrease there is a rapid fall during inflation, after which the density continues to decrease like $t^{-3/2}$ (orange dot-dashed line), since particle creation no longer takes place. Note that the density of the created particles is small compared to the total energy density in the universe (denoted in figure 2 by the red solid line) at any given time.

Figures 5--8 show the number density and energy density for the created particles with masses $m_2=10^3 \ \rm GeV$ (figures 5, 6) and $m_3=10^{-3} \ \rm GeV$ (figures 7, 8). There are two qualitative differences compared to the previous plots. Because the particles are lighter, the conditions $m t \ll 1$ and $m a \ll 1$ hold also for later times and the plots match the approximations (\ref{nden}) and (\ref{rrho2}) (green dashed line) all the way until the inflation epoch. Secondly, particle creation continues after inflation and stops only at $t\approx 1/m$,  which is confirmed by fitting the line (orange dot-dashed) going like $t^{-3/2}$.

\begin{figure}[h]
\centerline{\scalebox{1.5}{\includegraphics{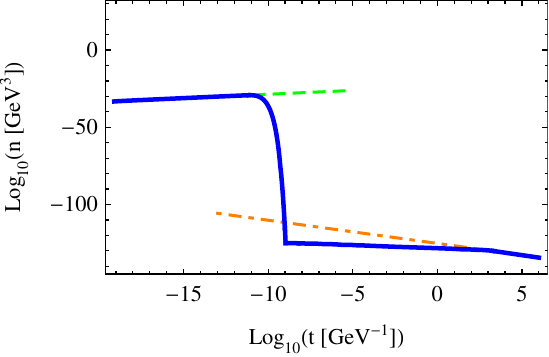}}}
\caption{\footnotesize{Same as figure 3, but for $m_3=10^{-3} \ \rm GeV$.}}
\end{figure}
\begin{figure}[h]
\centerline{\scalebox{1.5}{\includegraphics{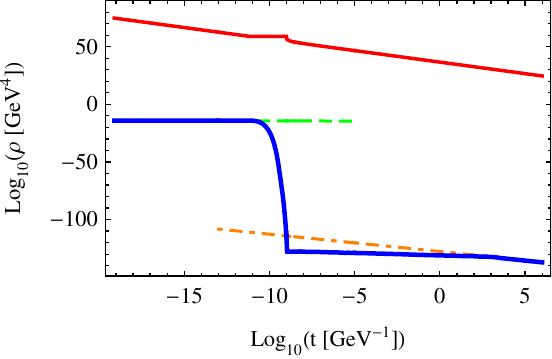}}}
\caption{\footnotesize{Same as figure 4, but for $m_3=10^{-3} \ \rm GeV$.}}
\end{figure}

We emphasize that although particle creation lasts arbitrarily long for light enough particles, their energy density is still negligible compared to the total energy density. For instance, if we wanted to have particles created today, we would need a particle of mass $m_{\rm light} \ll 10^{-42} \ \rm GeV$. From formula (\ref{rrho2}) the energy density is suppressed approximately by $m^4/a^4$. Taking into account also the huge decrease in energy density of the created particles during inflation, there is no way it can ever compete with the
total energy density of the universe.

\section{Zero mode}
Let us now consider the contribution of the zero mode. In this case all the formulae simplify significantly. Relation (\ref{omega}) becomes,
\bea
\omega_0(t) = m\, a(t)\ .
\eea
Equation (\ref{Bog}) for the Bogoliubov coefficient reduces to,
\bea\label{Bog0}
\ddot{\beta}_0 + \left(\frac{\dot{a}}{a} - \frac{\ddot{a}}{\dot{a}} + 2 i m \right)\dot{\beta}_0 - \frac{\dot{a}^2}{4 a^2}\beta_0 = 0\ ,
\eea
with the initial conditions,
\bea\label{initiial0}
\beta_0(t_0) = 0\ , \ \ \ \dot{\beta}_0(t_0) = \frac{\dot{a}(t_0)}{2\, a(t_0)}\ .
\eea
The equation for $u_0(t)$ is,
\bea\label{eq20}
\ddot u_0 + 3 \frac{\dot{a}}{a}\dot u_0 + \left(m^2+\frac{\ddot{a}}{a}+\frac{\dot{a}^2}{a^2}\right) u_0 = 0\ ,
\eea
where,
\bea\label{initiial00}
u_0(t_0) = \frac{1}{\sqrt{m \,a(t_0)}}\ , \ \ \ \dot{u}_0(t_0) = i \,\sqrt{\frac{m}{a(t_0)}}\ .
\eea

Once equations (\ref{Bog0}) and (\ref{eq20}) are solved, the particle number density, energy density and pressure can be calculated using the simple relations,
\bea\label{number0}
n(t) &=& \frac{|\beta_{0}|^2}{(2\pi a)^3}\ ,\\
\rho(t) &=& \frac{m |\beta_{0}|^2}{4\pi^3 a^3}  \ ,\\
p_{ij}(t) &=& \tilde{t}_{ij}\,\frac{m}{12\pi^3 a^3} \left(|\beta_{0}|^2-\frac{1}{2} m a |u_0|^2 +\frac{1}{2}\right).
\eea

It turns out that equation (\ref{Bog0}) with the initial conditions (\ref{initiial0}) can be solved exactly for simple choices of $a(t)$. For instance, in the pre-inflationary era with $a(t) = a_0 \sqrt{t}$, the solution for a particle of mass $m$ takes the form,
\bea\label{Leg}
& & \hspace{-7mm}\beta_0(t)= \frac{i}{2m} \,t^{\sfrac{1}{4}}\, t_0^{\sfrac{-5}{4}}\, e^{2 i m (t_0- t)}\bigg[U\left(\tfrac{5}{4},\tfrac{3}{2},2 i m t_0\right) \nonumber\\
 & & \times \, L_{-5/4}^{1/2}(2 i m t)-\,U\left(\tfrac{5}{4},\tfrac{3}{2},2 i m t\right) L_{-5/4}^{1/2}(2 i m t_0)\bigg]\nonumber\\
& & \times  \bigg[4\,
   U\left(\tfrac{5}{4},\tfrac{3}{2},2 i m t_0\right) L_{-9/4}^{3/2}(2 i m t_0)\nonumber\\
& &\hspace{6mm} -5 \,U\left(\tfrac{9}{4},\tfrac{5}{2},2 i m
   t_0\right) L_{-5/4}^{1/2}(2 i m t_0)\bigg]^{-1},
\eea
where $U(x,y,z)$ is the confluent hypergeometric function of the second kind, and $L^k_n(x)$ is the associated Laguerre polynomial.

In the limit $m t \ll 1$ the dependence of the Bogoliubov coefficients in formula (\ref{Leg}) on $t$ and $t_0$ is $\beta_0 \sim (t/t_0)^{\sfrac{1}{4}}$, which translates to,
\bea\label{cond}
n(t, t_0) \sim \rho(t, t_0) \sim \frac{1}{t\,\sqrt{t_0}}\ .
\eea
Thus, setting initial conditions at a small enough $t_0$, one can have an arbitrarily large zero mode particle production at any given instance of time. As mentioned earlier, this situation doesn't occur for the nonzero modes, for which $\dot{\beta}_\lambda(t_0)$ is independent of $t_0$ for $m a \ll 1$.
However, since we don't know the physics before the Planck time, choosing $t_0 < t_p$ is not justified.
In the subsequent calculation we assume $t_0 = t_p$, just as in the case of the nonzero modes.

It turns out that also equation (\ref{eq20}) possesses an analytical solution for $a(t)=a_0\sqrt{t}$, given by,
\bea\label{Leg2}
& &\hspace{-3mm} u_0(t) = \frac{\pi}{4(m a_0 \sqrt{t})^{1/2}} \Bigg\{Y_{1/4}(m t) \bigg[-2 m
   t_0 J_{-3/4}(m t_0)\nonumber\\
   & &
   \hspace{12mm}+\left(1+2 i m
   t_0\right) J_{1/4}(m
   t_0)\bigg]+J_{1/4}(m t) \nonumber\\
   & & \ \ \ \times \bigg[2 m t_0
   Y_{-3/4}(m t_0)
   -\left(1+2 i m
   t_0\right) Y_{1/4}(m t_0)\bigg]\Bigg\}\ ,\nonumber\\
\eea
where $J_\alpha(x)$ is the Bessel function of the first kind and $Y_\alpha(x)$ is the Neumann function.

\begin{figure}[t]
\centerline{\scalebox{1.5}{\includegraphics{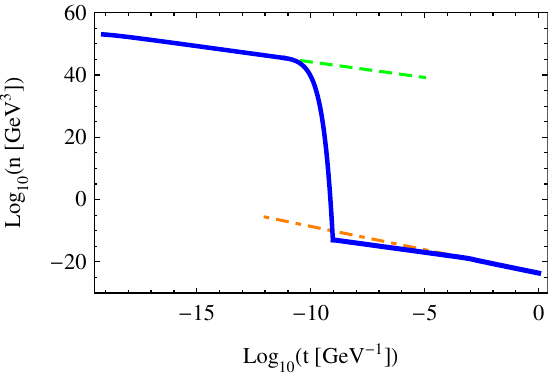}}}
\caption{\footnotesize{Number density of the particles created through the zero mode for $m=10^3 \ \rm GeV$ (blue solid line), assuming $t_0=t_p$. The green dashed line corresponds to the $m t \ll 1$ approximation given by equation (\ref{cond}). The orange dot-dashed line denotes the $m t \gg 1$ approximation given by equation (\ref{approx}).}}
\end{figure}
\begin{figure}[t]
\centerline{\scalebox{1.5}{\includegraphics{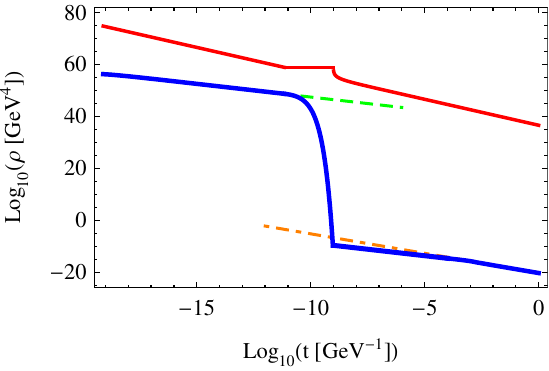}}}
\caption{\footnotesize{Energy density of the particles created through the zero mode for $m=10^3 \ \rm GeV$ (blue solid line), assuming $t_0=t_p$. The green dashed line corresponds to the $m t \ll 1$ approximation given by equation (\ref{cond}). The orange dot-dashed line corresponds to the $m t \gg 1$ approximation given by equation (\ref{approx}). The red solid line denotes the total energy density of the universe.}}
\end{figure}

Figure 9 shows the number density and figure 10 presents the energy density for particles of mass $m=10^3 \ \rm GeV$ created through the zero mode, assuming a full evolution of $a(t)$ given in table I, with the initial conditions set at $t_0=t_p$. In the region $m t \ll 1$ both the number density and energy density decrease according to (\ref{cond}), which is confirmed by fitting the line (green dashed) going like $1/t$. Because of choosing a small mass of the created particles, the fit matches the plot until the beginning of inflation, during which both the number density and energy density fall rapidly by many orders of magnitude.
The orange dot-dashed line corresponds to,
\bea\label{approx}
n(t) \sim \rho(t) \sim \frac{1}{t^{\sfrac{3}{2}}} \sim \frac{1}{a^3} \ ,
\eea
i.e., no particle creation. It fits well the plot for $t \gtrsim 1/m$, which implies that for the zero mode particle creation also stops at $t \approx 1/m$. The energy density of the created particles is again small compared to the total energy density in the universe.\footnote{We note, however, that if $\rho(t) \ll M_p^4$ for a range of $t < t_p$ and the equations in this paper are still valid in that region, by assuming a small enough $t_0$ the energy density of the particles created through the zero mode might in general be large enough to compete with ordinary matter and radiation at some instance of time in the evolution of the universe.}

\begin{figure}[t]
\centerline{\scalebox{1.5}{\includegraphics{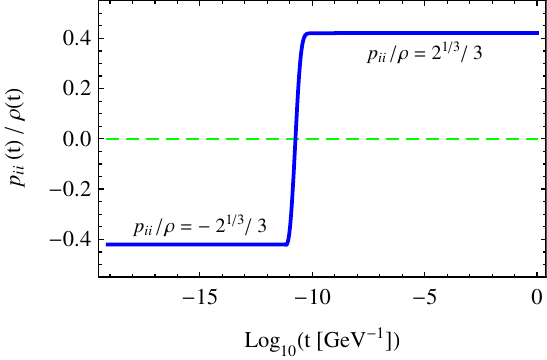}}}
\caption{\footnotesize{Ratio of the pressure $p_{ii}(t)$ and the energy density $\rho(t)$ of the particles created through the zero mode for $m=10^3 \ \rm GeV$ assuming the initial condition at $t_0=t_p$.}}
\end{figure}

It is interesting to analyze the equation of state for the matter created through the zero mode, as it has significantly different properties than in the nonzero mode case.
Let us for a moment assume that the entire evolution proceeds according to the pre-inflationary expansion $a(t) = a_{p0} \sqrt{t}$. An analysis of formulae (\ref{Leg}) and (\ref{Leg2}) shows that for such a scenario,
\bea\label{state}
p_{ij}(t) =
\begin{cases}
-\tilde{t}_{ij}\,\frac{1}{3}\rho(t) & \ \ \ \  {\rm for} \ \ \ m t \ll 1\ ,\\
\ \ \tilde{t}_{ij}\,\frac{1}{3}\rho(t) & \ \ \ \   {\rm for} \ \ \ m t \gg 1\label{state2}\ .
\end{cases}
\eea
Since $\tilde{t}_{ii} = \sqrt[3]{2}$ and $\tilde{t}_{ij} = 1/\sqrt[3]{4}$ for $i\ne j$, the created matter would have a ``quasi-accelerating'' equation of state for times $t \lesssim 1/m$. This picture, however, is significantly modified when we take into account inflation.

Figure 11 shows the the ratio $p_{ii}(t)/\rho(t)$ for the zero mode production of particles with mass $m=10^3 \ \rm GeV$ including the full evolution of the three-torus universe from table I. As expected from equation (\ref{state}), for the pre-inflationary epoch it has the form $p_{ii}(t)/\rho(t)=-\sqrt[3]{2}/3$ . However, it doesn't stay like this until $t \approx 1/m$ -- inflation itself changes the equation of state to $p_{ii}(t) = \rho(t) \sqrt[3]{2}/3$, and it remains this way for the rest of the evolution. This behavior seems generic for all masses of the created particles. An interesting fact is that for much shorter inflation times one could end up with an equation of state with $p_{ii}(t)/\rho(t)$ anywhere between $-\sqrt[3]2/3$ and $\sqrt[3]2/3$ at the end of inflation. However, such short inflation periods are not realistic.

\section{Conclusions}
In the present work we calculated the rate for gravitational particle creation in a three-torus universe.
We performed the calculation assuming the metric on the three-torus for which the shape moduli are stable. We adopted the current size of the universe of ten Hubble radii and estimated its full evolution in time. We argued that the Casimir energies might dominate the total energy density before inflation, resulting in the expansion of the universe like in the radiation era.
We then showed that there are two types of contributions to the particle production rate -- one from the nonzero modes on a torus, and the other coming from the zero mode.

The nonzero mode contribution is interesting because at early times it is characterized by a ``quasi-vacuum-like'' equation of state for the produced matter. The particle production itself continues until $t \approx 1/m$, after which it essentially stops. It turns out that the energy density of particles created through the nonzero modes is tiny compared to the total energy density in the universe at all stages of its evolution.

The production through the zero mode is more unique. The corresponding energy density of the created particles is very sensitive to initial conditions.
If we set those at the Planck time, the energy density of the particles is again small with respect to the total energy density of the universe. We showed that although the particles produced through the zero mode have a ``quasi-accelerating'' equation of state  before inflation, they are described by a ``quasi-radiative'' equation afterwards.

\subsection*{Acknowledgment}
The author is extremely grateful to Mark Wise for stimulating discussions and helpful comments.
The work was supported in part by the U.S. Department of Energy under contract No. DE-FG02-92ER40701.



\begin{thebibliography}{99}

\bibitem{Komatsu:2010fb}
  E.~Komatsu {\it et al.}  [WMAP Collaboration],
  \textit{Seven-year Wilkinson Microwave Anisotropy Probe (WMAP) observations: Cosmological interpretation},
  Astrophys.\ J.\ Suppl.\  {\bf 192}, 18 (2011)
  [arXiv:1001.4538 [astro-ph.CO]].

\bibitem{Glen}
  P.~M.~Vaudrevange, G.~D.~Starkman, N.~J.~Cornish and D.~N.~Spergel,
  \textit{Constraints on the topology of the universe: Extension to general geometries},
  arXiv:1206.2939 [astro-ph.CO].

\bibitem{ArkaniHamed:2007gg}
  N.~Arkani-Hamed, S.~Dubovsky, A.~Nicolis and G.~Villadoro,
  \textit{Quantum horizons of the standard model landscape},
  JHEP {\bf 0706}, 078 (2007)
  [hep-th/0703067].

\bibitem{AFW}
  J.~M.~Arnold, B.~Fornal and M.~B.~Wise,
  \textit{Standard model vacua for two-dimensional compactifications},
  JHEP {\bf 1012}, 083 (2010)
  [arXiv:1010.4302 [hep-th]].

\bibitem{FW}
  B.~Fornal and M.~B.~Wise,
  \textit{Standard model with compactified spatial dimensions},
  JHEP {\bf 1107}, 086 (2011)
  [arXiv:1106.0890 [hep-th]].

\bibitem{Linde}
  A.~D.~Linde,
  \textit{Creation of a compact topologically nontrivial inflationary universe},
  JCAP {\bf 0410}, 004 (2004)
  [hep-th/0408164].

\bibitem{Zeldov}
  Y.~B.~Zel'dovich, A.~A.~Starobinsky,
  \textit{Quantum creation of a universe in a nontrivial topology},
  Sov.\ Astron.\ Lett. {\bf 10}, 135 (1984).

\bibitem{Parker1}
  L.~Parker,
  \textit{Particle creation in expanding universes},
  Phys.\ Rev.\ Lett.\  {\bf 21}, 562 (1968).

\vspace{0.8mm}

\bibitem{Parker2}
  L.~Parker,
  \textit{Quantized fields and particle creation in expanding universes. 1.},
  Phys.\ Rev.\  {\bf 183}, 1057 (1969).

\bibitem{Parker3}
  L.~Parker,
  \textit{Quantized fields and particle creation in expanding universes. 2.},
  Phys.\ Rev.\ D {\bf 3}, 346 (1971)
  [Erratum-ibid.\ D {\bf 3}, 2546 (1971)].

\bibitem{Zeldovich}
  Y.~B.~Zel'dovich, A.~A.~Starobinsky,
  \textit{Particle production and vacuum polarization in an anisotropic gravitational field},
  Sov. Phys. JETP {\bf 34}, 1159 (1972).

\bibitem{Starob}
  S.~G.~Mamaev, V.~M.~Mostepanenko and A.~A.~Starobinsky,
  \textit{Particle creation from the vacuum near a homogeneous isotropic singularity},
  Sov. Phys. JETP {\bf 43}, 823 (1976).

\bibitem{Grib}
  A.~A.~Grib, S.~G.~Mamayev and V.~M.~Mostepanenko,
  \textit{Particle creation from vacuum in homogeneous isotropic models of the universe},
  Gen.\ Relativ.\ Gravit. {\bf 6}, 535 (1976).

\bibitem{Chung}
  D.~J.~H.~Chung, P.~Crotty, E.~W.~Kolb and A.~Riotto,
  \textit{Gravitational production of superheavy dark matter},
  Phys.\ Rev.\ D {\bf 64}, 043503 (2001)
  [hep-ph/0104100].

\bibitem{Berger}
  B.~K.~Berger,
  \textit{Scalar particle creation in an anisotropic universe},
  Phys.\ Rev.\ D {\bf 12}, 368 (1975).

\bibitem{Aneesh}
  G.~Aslanyan and A.~V.~Manohar,
  \textit{The topology and size of the universe from the cosmic microwave background},
  arXiv:1104.0015 [astro-ph.CO].


\end{thebibliography}
\end{document}